# ARTICLE INFORMATION

**Article title**

Full-field laser-Doppler-vibrometry FRF dataset for a four-bolt aluminum plate under bolt-torque variation

**Authors**

Berkay Kullukcu*, Dina Hannebauer

**Affiliations**

FG Machine Dynamics and Acoustics, TH Wildau, Hochschulring 1, 15745 Wildau, Brandenburg, Germany

**Corresponding author's email address and Twitter handle**

berkay.kullukcu@th-wildau.de / @kullukcu_berkay

**Abstract**

The dataset described in this article was constructed using the results of laser Doppler vibrometry experiments conducted on an aluminum plate having four bolts. The database captures the variation in the vibration characteristics of the plate as a function of changes in tightening torque on the bolts. In total, there were 18 different states for the tightening torque, ranging from a situation where all bolts are tightened to all bolts being loose. Other possible combinations include states with varying torque applied to just one bolt, and those involving multiple loose bolts. Data consist of 1,836 files exported from Polytec PSV-500, categorized as amplitude and phase data, for frequency ranges spanning 1 Hz to 51,200 Hz at intervals of 1 Hz. The identified zones were analyzed at 51 scan points for all states of torque. Supporting documents included specifications of the test plate, scan point and torque state information, selected resonance peaks tables, measured frequencies for seven resonance groups, correlation values for all FRF, visualizations, and code scripts.

# SPECIFICATIONS TABLE

| | |
|---|---|
| **Subject** | Engineering & Materials science |
| **Specific subject area** | Structural dynamics, vibro-acoustic measurements, and laser-Doppler-vibrometry data for bolted-joint diagnostics. |
| **Type of data** | Pointwise FRF exports (TXT), documentation and processed tables (CSV), images (PNG), scripts (PY), metadata (JSON/JSON-LD/CFF/YAML), text documentation (MD/TXT), packaged dataset (ZIP). |
| **Data collection** | A Polytec PSV-500 scanning laser Doppler vibrometer measured pointwise H1 amplitude and phase of displacement-to-force for a four-bolt aluminum plate subjected to excitation |
| | using an automatic impact hammer. Soft sponges were placed below the plate to simulate free-free boundary conditions. The selected 18 torque states were exported: 1 “all tight” state, 1 “all loose” state, 4 cases where only one bolt is tightened to 5Nm, 4 cases where only one bolt is loosened to 0Nm, 4 cases of three loose bolts and one tight bolt, and 4 cases of two loose bolts. |
| **Data source location** | Institution: TH Wildau; City: Wildau; Region: Brandenburg; Country: Germany. |
| **Data accessibility** | Repository name: Zenodo<br>Dataset name: Full-field laser-Doppler-vibrometry frequency-response-function data for a four-bolt aluminum plate under bolt-torque variation<br>Data identification number: 10.5281/zenodo.20038951<br>Direct URL to data: https://zenodo.org/records/20038951 |
| **Related research article** | B. Kullukcu, R. Pianowski, M.S. Özer, E. Altinsoy, D. Hannebauer, Screening bolt loosening in a four-bolt plate with global FRF correlation and local FRAC maps from full-field laser Doppler vibrometry, manuscript submitted for publication, 2026 [1]. |

# VALUE OF THE DATA

- The dataset [2] contains 18 selected bolt-torque combinations, which are available as the pointwise H1 amplitude and phase data sets, including the four single-bolt series at 10, 5, and 0 Nm and four combined multi-loose setups.
- For researchers investigating the FRF-correlation, modal assurance criterion, or FRF-shape damage localization techniques, the 51-point common subset and 1 Hz spectral resolution should be adequate. These parameters will provide a good platform for benchmarking alternative similarity measures or selection criteria [3–6].
- Instructors interested in vibration-based structural health monitoring, modal testing, or bolted joint dynamics may want to utilize this data set as a learning tool. All torque levels, scan point coordinates, and pre-processed data tables are included.
- Data tables are partitioned into frequency candidate detection and retention, frequency tracking, per-case statistics, and per-frequency family statistics sections. Such organization allows for easy evaluation of multiple selection or screening procedures without having to repeat the entire analysis cycle from the beginning.
- Intermediate 5 Nm cases provide an opportunity to investigate dose-response relationships across three torque levels for each bolt. Although such aspects were not fully discussed in the original study, now there is an opportunity to work with raw data and conduct analysis.

# BACKGROUND

The short communication [1] applies resonance-group screening to bolt-torque variation in a four-bolt plate and is supported by the data package [2]. The accompanying research article makes use of full-field laser Doppler vibrometry to compare an all-tight baseline against four single-bolt loose cases using both amplitude-only and phase-inclusive FRF correlation metrics along with local FRAC maps. The data article provides supplementary context by decoupling the data products from its interpretation, and by making available the further torque configurations contained in the raw exports but not included in the screening.

Bolted joints are an established source of uncertainty in structural dynamics because preload controls contact stiffness, frictional dissipation, microslip, and local boundary conditions. Reviews and recent model-identification studies show that changes in bolt condition can shift resonance frequencies and reshape frequency-response functions, making well-documented benchmark datasets useful for testing damage indicators [7–10]. Full-field laser Doppler vibrometry is relevant in this context because it captures the spatial redistribution of vibration over the measured surface rather than only at a few discrete sensor locations; continuous-scan and full-field LDV studies have demonstrated this advantage for modal identification and damage-sensitive response-shape analysis [11–13]. The processed tables in this dataset therefore connect the raw H1 amplitude and phase exports to established correlation-based tools such as MAC, FRAC, and FRF-shape damage localization [3–6].

# DATA DESCRIPTION

The dataset consists of root documentation/metadata file(s) and six main folders: 00_raw_exports (H1 amplitude and phase pointwise TXT files for the 18 torque states), 01_documentation, 02_processed_tables, 03_figures, 04_scripts, and metadata.

There are six graphs presented in the paper. Fig. 1 depicts the test setup. Fig. 2 represents the in-plane spatial mapping of the 51 LDV points common across all 18 states of torque. The locations of the bolts are shown by star symbols. Fig. 3 and 4 present the median of spatial-RMS spectra according to the torque-state classes for 1-12 kHz and 1-51 kHz, respectively. Fig. 5 describes the selected candidate peaks of the all-tight spatial-RMS spectrum. Fig. 6 provides example local FRF-correlation maps included in the processed-data folder.

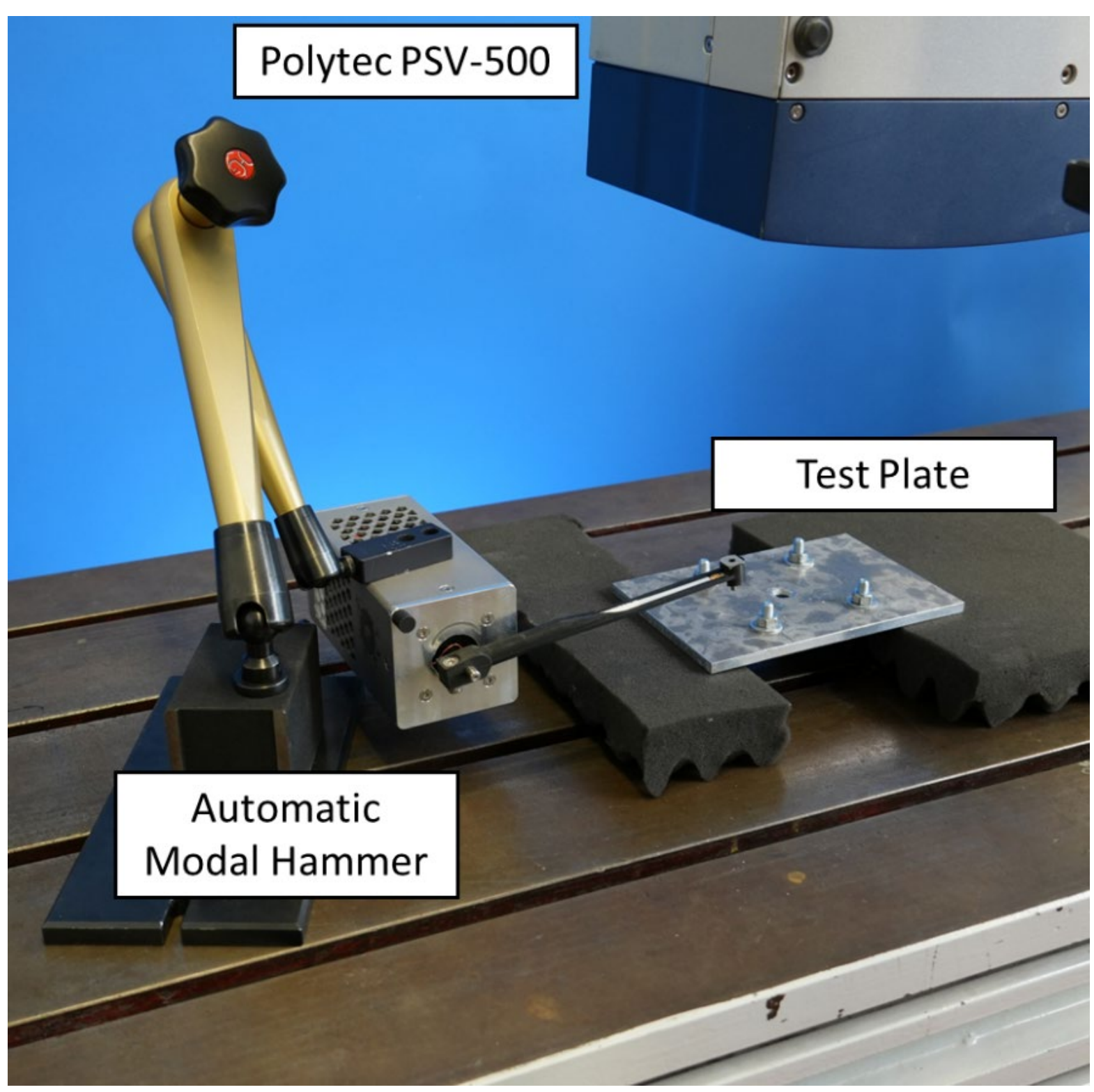

**Fig. 1.** Experimental setup showing the Polytec PSV-500 scanning laser Doppler vibrometer, the four-bolt aluminum plate on soft supports, and the automatic modal hammer used for repeated excitation

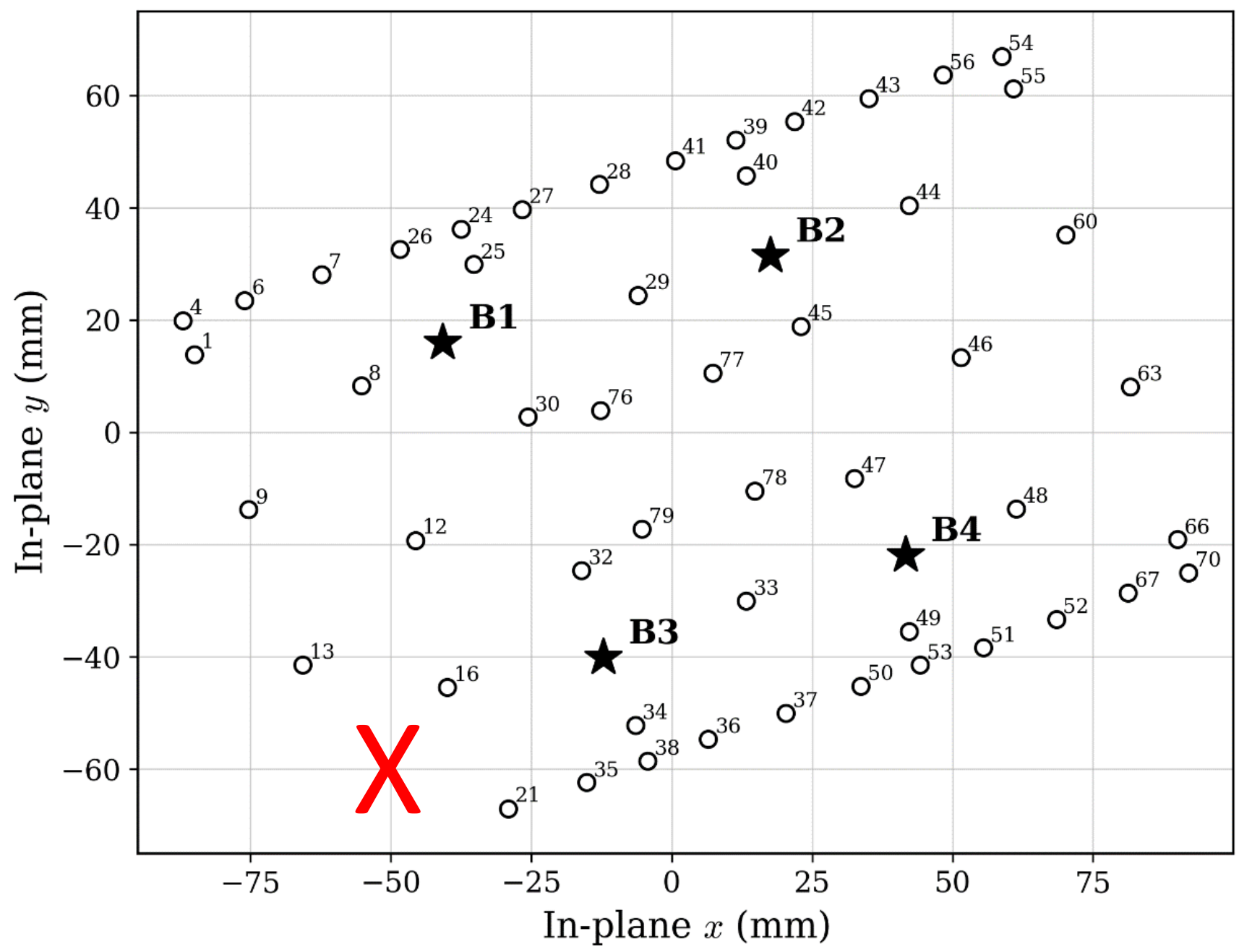

**Fig. 2.** Spatial layout of the 51 LDV points (circles) common to all 18 torque states, projected onto the plate plane. Bolt centers B1–B4 are marked with stars. Numbers next to each point are the indices used in the raw export filenames. The red cross marks the impact point of the hammer.

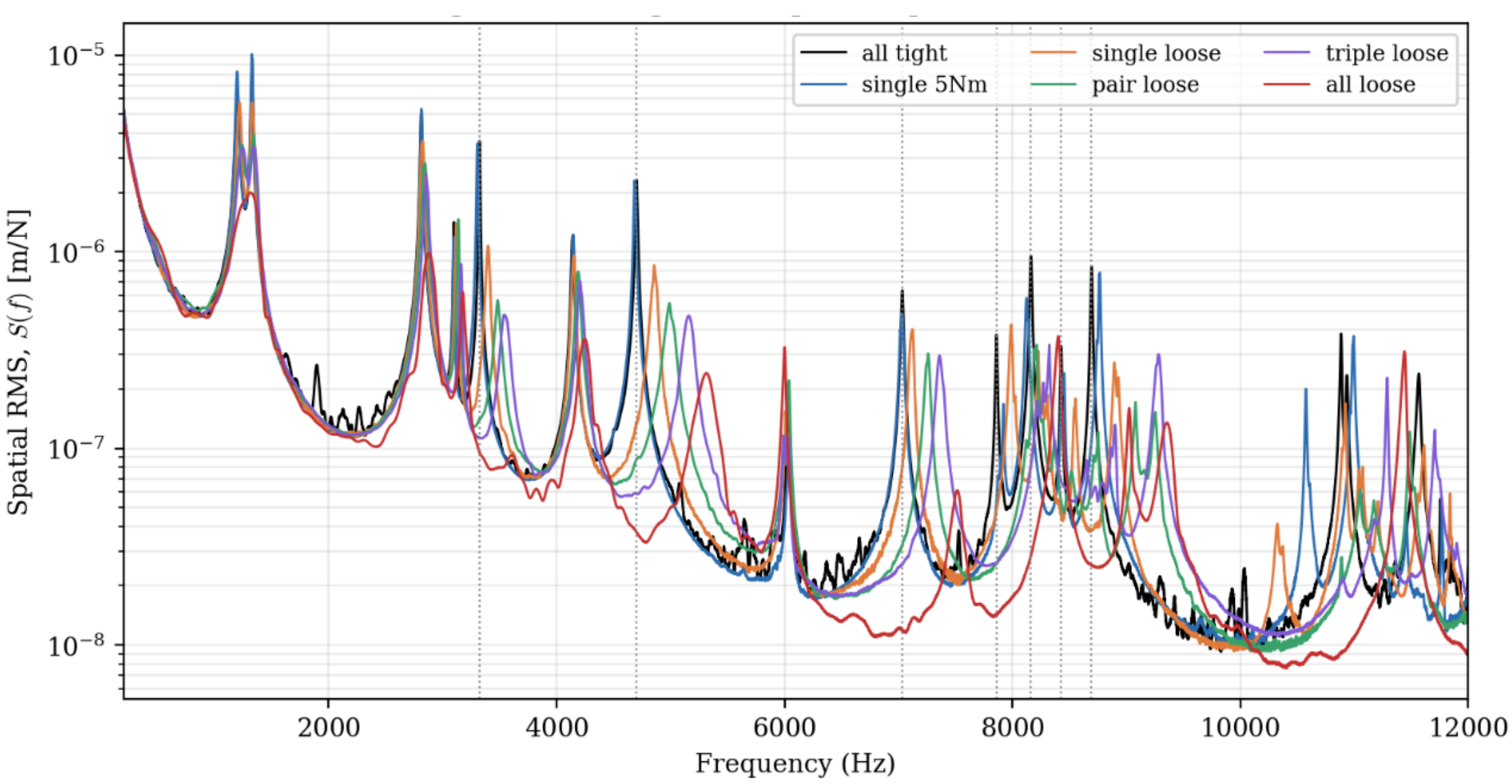

**Fig. 3.** Spatial-RMS spectra across the 18 torque states (median per torque-state class) from 1 Hz to 12 kHz. The retained resonance groups are marked by vertical dotted lines.

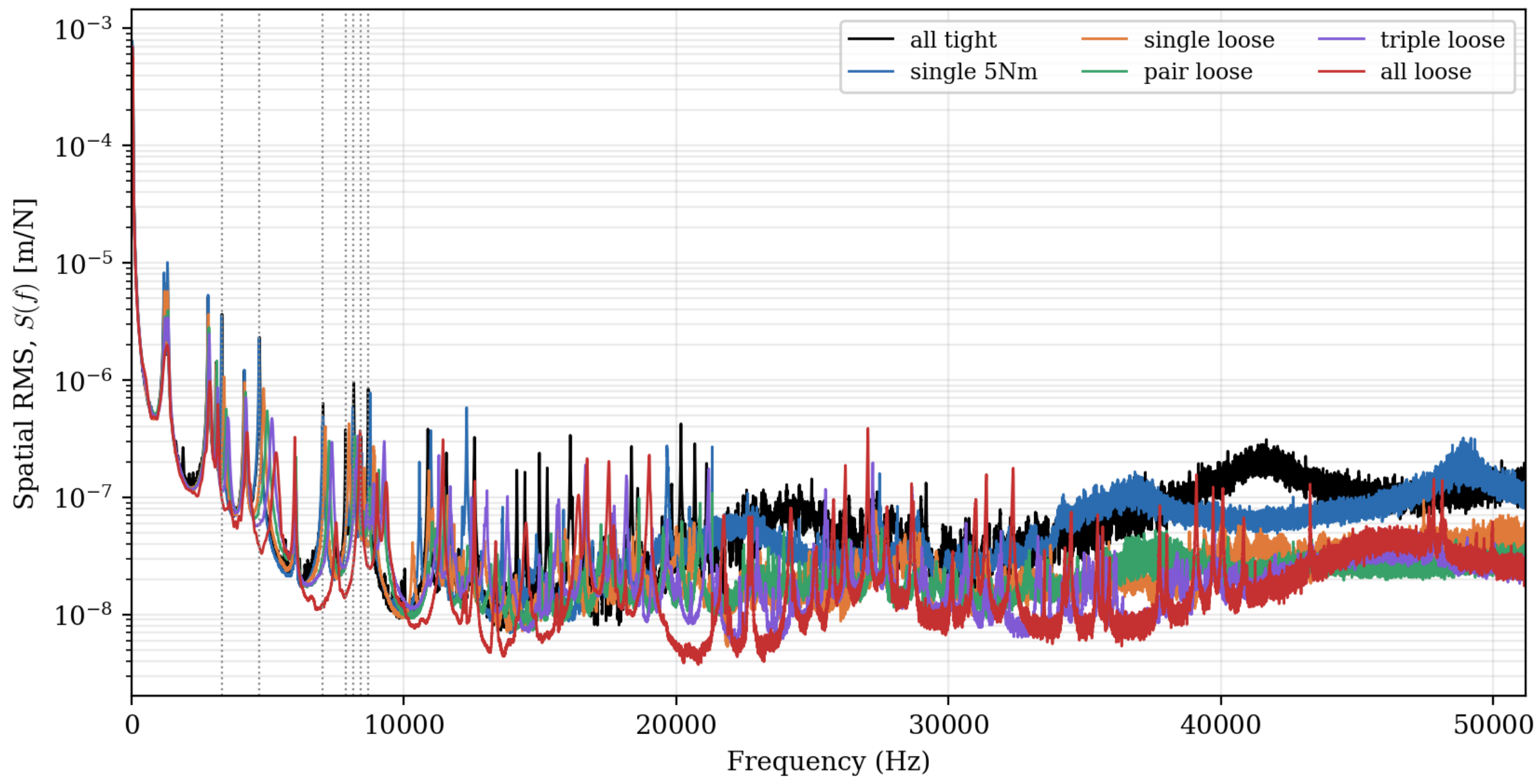

**Fig. 4.** Spatial-RMS spectra across the 18 torque states (median per torque-state class) from 1 Hz to 51 kHz.

Table 1 lists the top-level folders. Table 2 summarizes the processed tabular datasets that a secondary user is most likely to inspect first. Table 3 is a compact variable dictionary for the main processed variables. Table 4 lists the eighteen torque configurations represented in the raw exports. Table 5 summarizes the dominant file types stored in the package.

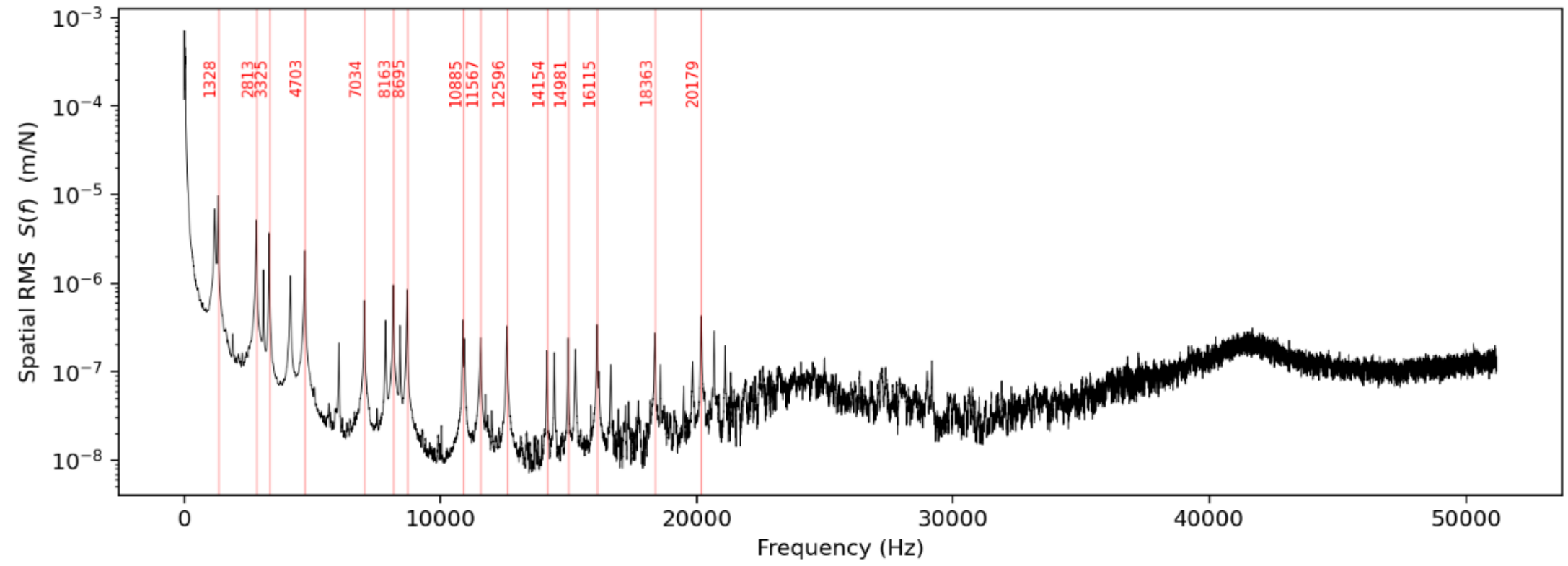

**Fig. 5.** All-tight reference spatial-RMS spectrum with the picked candidate resonance peaks marked by red vertical lines. The retained resonance groups are documented numerically in the retention-audit and adaptive-tracking-window tables.

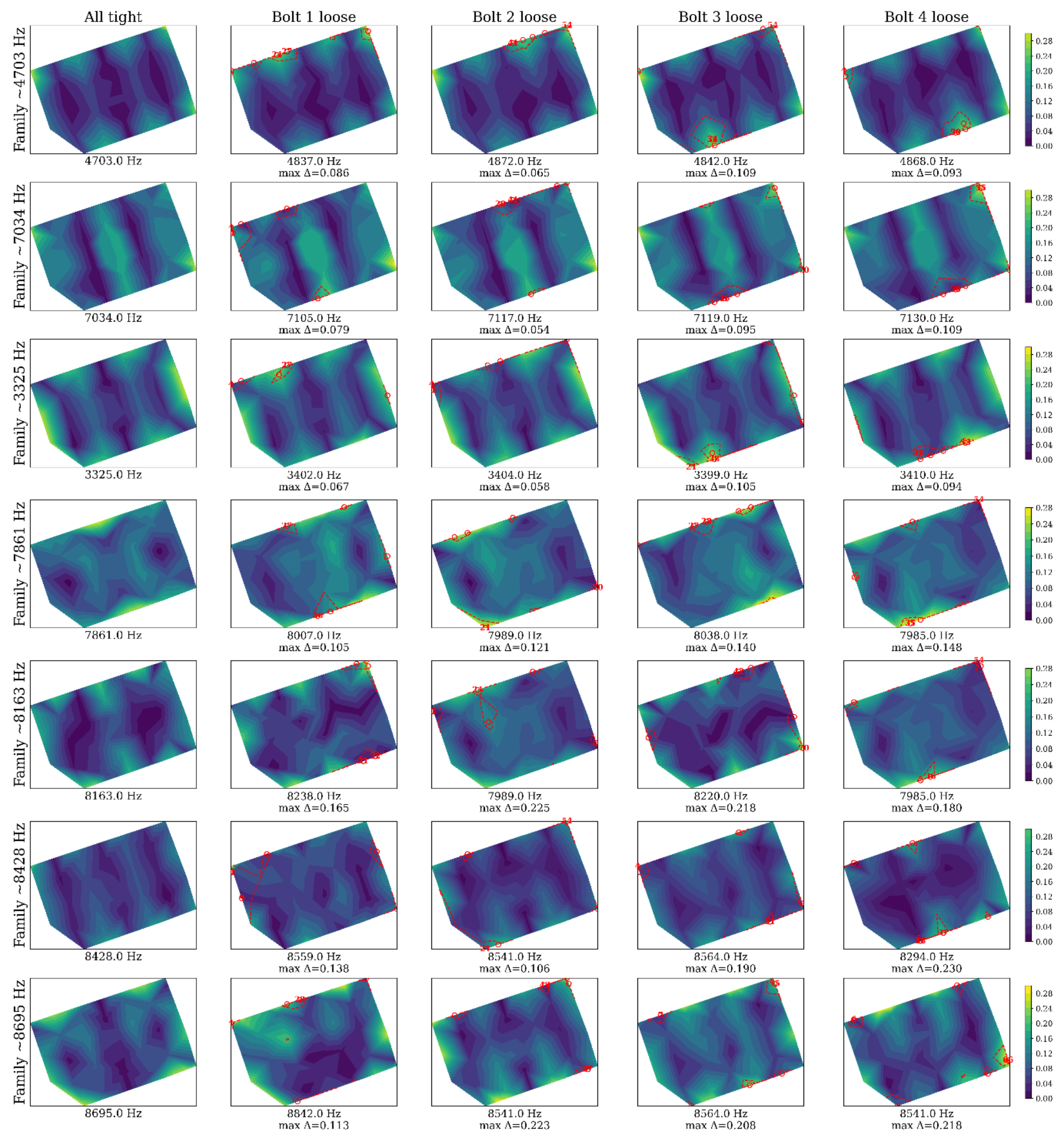

**Fig. 6.** Local FRF-correlation sensitivity maps for the retained frequency groups (families), comparing the all-tight baseline with the four single-bolt-loose cases.

**Table 1.** Top-level folder layout.

| top_level_folder | purpose | n_files |
|---|---|---|
| 00_raw_exports | Pointwise amplitude and phase H1 FRF TXT exports for the 18 torque configurations. | 1836 |
| 01_documentation | Setup photograph, raw-export format note, scan-point coordinate list, torque-state metadata, and folder README. | 5 |
| 02_processed_tables | Seven processed CSV tables plus folder README. | 8 |
| 03_figures | Six PNG figures used in the data article plus folder README. | 7 |

| 04_scripts | Six Python scripts plus folder README for rebuilding processed tables, figures, and optional HDF5 export. | 7 |
|---|---|---|
| metadata | DataCite and Schema.org machine-readable metadata files. | 2 |

**Table 2.** Key processed tabular datasets.

| file_name | content | rows | columns | key_variables |
|---|---|---|---|---|
| candidate_resonance_groups.csv | Discovery-set candidate peaks derived from the all-tight spatial-RMS spectrum. | 15 | 4 | $f_{\text{peak}}$; $\log_{10}(p)$; $\text{BW}_{\text{HP}}$; $\Delta f_{\text{excl}}$ |
| resonance_group_selection_audit.csv | Audit table documenting why candidates were retained or excluded. | 15 | 8 | $\text{rank}(p)$ $f_{\text{peak}}$; $r$; $\Delta f_{\text{win}}$; $\text{rule}_{\text{ret}}$ |
| tracked_frequencies.csv | Tracked frequencies of the seven retained resonance groups across the all-tight and four single-bolt-loose cases. | 7 | 6 | $f_{\text{family}}$; $f_{\text{tight}}$; $f_{\text{bolt1}}, \ldots, f_{\text{bolt4}}$ |
| per_case_metrics.csv | Case-wise amplitude- | 28 | 7 | $g$; |

| | only, phase-aware, and phase-only global dissimilarity metrics. | | | $b$<br>$f_{\text{base}}$;<br>$f_{\text{case}}$;<br>1-MAC$_a$;<br>1-CMAC;<br>1-CMAC$_{\text{phase}}$ |
|---|---|---|---|---|
| family_summary.csv | Family-wise mean, 95% exhaustive-bootstrap interval, extrema, and compactness summary metrics. | 7 | 17 | $f_{\text{family}}$;<br>$\overline{\text{1-MAC}_a}$;<br>$\overline{\text{1-CMAC}}$;<br>$\overline{C_5}$;<br>$\overline{R_5}$ |
| dose_response.csv | Per-bolt 10/5/0 Nm dose-response of the global metrics for every retained resonance group. | 84 | 7 | $g$;<br>$b$;<br>$T$;<br>$f_{\text{tracked}}$;<br>1-MAC$_a$;<br>1-CMAC;<br>1-CMAC$_{\text{phase}}$ |
| adaptive_tracking_windows.csv | Tracking-window definitions for retained resonance groups. | 7 | 6 | $f_{\text{ret}}$;<br>BW$_{\text{HP}}$;<br>$\Delta f_{\text{cand}}$;<br>$\Delta f_{\text{win}}$ |

**Table 3.** Variable dictionary for the main processed variables.

| variable | units | description |
|---|---|---|
| $f_{\text{peak}}$ | Hz | Center frequency of a discovery-set candidate peak. |
| $\log_{10}(p)$ | log10 scale | Peak prominence on the smoothed spatial-RMS spectrum. |
| $\text{BW}_{\text{HP}}$ | Hz | Half-power bandwidth of a candidate or retained resonance group. |
| $r$ | Boolean | True if the candidate passed the unique-trackability rule. |
| $\Delta f_{\text{win}}$ | Hz | Rounded half-width of the tracking window assigned to a retained group. |
| $f_{\text{tight}}$ | Hz | Tracked resonance frequency in the all-tight baseline case. |
| $f_{\text{bolt1}}, \ldots, f_{\text{bolt4}}$ | Hz | Tracked resonance frequencies in the four single-bolt-loose cases. |
| $1\text{-MAC}_a$ | dimensionless | Case-wise amplitude-only global dissimilarity. |
| 1-CMAC | dimensionless | Case-wise phase-aware global dissimilarity. |
| $1\text{-CMAC}_{\text{phase}}$ | dimensionless | Case-wise phase-only complex similarity (CMAC of unit-modulus phase pattern). |
| $T$ | Nm | Bolt torque level (10, 5, or 0) for the dose-response analysis. |
| $\overline{C_5}$ | dimensionless | Family-wise mean compactness score C5 for local FRAC-deficit maps. |

**Table 4.** Eighteen torque configurations exported in 00_raw_exports.

| case_name | human-readable label | bolt torques (Nm) for bolts 1, 2, 3, 4 |
|---|---|---|
| 1234_10Nm | all tight (reference) | 10, 10, 10, 10 |
| 1234Lose | all loose | 0, 0, 0, 0 |

| 234_10Nm_1Lose | bolt 1 loose | 0, 10, 10, 10 |
|---|---|---|
| 234_10Nm_1_5Nm | bolt 1 at 5 Nm | 5, 10, 10, 10 |
| 134_10Nm_2Lose | bolt 2 loose | 10, 0, 10, 10 |
| 134_10Nm_2_5Nm | bolt 2 at 5 Nm | 10, 5, 10, 10 |
| 124_10Nm_3Lose | bolt 3 loose | 10, 10, 0, 10 |
| 124_10Nm_3_5Nm | bolt 3 at 5 Nm | 10, 10, 5, 10 |
| 123_10Nm_4Lose | bolt 4 loose | 10, 10, 10, 0 |
| 123_10Nm_4_5Nm | bolt 4 at 5 Nm | 10, 10, 10, 5 |
| 1_10Nm_234Lose | bolts 2, 3, 4 loose | 10, 0, 0, 0 |
| 2_10Nm_134Lose | bolts 1, 3, 4 loose | 0, 10, 0, 0 |
| 3_10Nm_124Lose | bolts 1, 2, 4 loose | 0, 0, 10, 0 |
| 4_10Nm_123Lose | bolts 1, 2, 3 loose | 0, 0, 0, 10 |
| 13_10Nm_24Lose | bolts 2, 4 loose (diagonal pair) | 10, 0, 10, 0 |
| 24_10Nm_13Lose | bolts 1, 3 loose (diagonal pair) | 0, 10, 0, 10 |
| 14_10Nm_23Lose | bolts 2, 3 loose (adjacent pair) | 10, 0, 0, 10 |
| 23_10Nm_14Lose | bolts 1, 4 loose (adjacent pair) | 0, 10, 10, 0 |

**Table 5.** Dominant file types in the package.

| extension | description | n_files |
|---|---|---|
| txt | Pointwise H1 amplitude and phase FRF exports listed in CHECKSUMS.sha256. | 1836 |
| csv | Documentation codebooks and processed analysis tables. | 9 |
| png | Setup photograph and data-article figures. | 7 |
| py | Analysis, figure-rebuild, and HDF5-export scripts. | 6 |

# EXPERIMENTAL DESIGN, MATERIALS AND METHODS

Test specimen is an aluminum plate with four diagonally placed M6 bolts and has an additional M6 hole in its center, which was not used during the tests. The specimen has the dimensions of 100x150x7 [mm]. Specimen was placed on soft sponges that simulated free-free conditions, while vibration excitation was performed with the help of an automated modal hammer. Displacement-to-force H1 frequency-response functions were recorded using a Polytec PSV-500 scanning laser Doppler vibrometer. The scanning points used for analysis were nominated as specified in the respective scan-point-layout and torque state metadata files. Torque wrench was used to fasten bolts 1 through 4, which were arranged on the plate according to the sketch and coordinates in Fig. 2.

The raw TXT exports contain pointwise H1 displacement-to-force frequency-response functions. Amplitude files report the magnitude of the receptance FRF, while phase files report the corresponding phase angle. The amplitude header in the .txt exports identify the signal as "FFT - Vib & Ref1 H1 Weg / Kraft – Amplitude" (displacement divided by force) and gives the amplitude unit as m / N. For each LDV scan point, the exported amplitude can be interpreted as given in Eq. 1:

$$|H_1(f)| = \left|\frac{x(f)}{F(f)}\right|, \text{ (Eq. 1)}$$

with units given in Eq. 2:

$$[H_1(f)] = \mathrm{m/N} \text{ (Eq. 2)}$$

Here, $x(f)$ is the displacement response at frequency $f$, and $F(f)$ is the reference force input. The phase TXT exports correspond to the phase of the same H1 displacement-to-force FRF. The phase header identifies the signal as "FFT - Vib & Ref1 H1 Weg / Kraft – Phase" (displacement divided by force), with phase values reported in degrees. Therefore, when amplitude and phase are recombined into a complex FRF for scan point $i$, the phase must be converted from degrees to radians (Eq. 3):

$$H_i(f) = A_i(f)\exp\left(j\frac{\pi}{180}\phi_i(f)\right), \text{ (Eq. 3)}$$

where $A_i(f)$ is the exported amplitude in $\mathrm{m/N}$ and $\phi_i(f)$ is the exported phase in degrees at LDV point $i$.

The main analysis script intersects the available point indices for the 18 torque states, preserving only the 51 coordinate points present in each state. The script generates the spatial root mean square spectrum of the all-tight case, discovers candidate peaks through smoothed log peak picking, analyses bandwidth and spacing information, and performs the unique trackability criterion described in resonance_group_selection_audit.csv. Tracking windows for the retained groups utilize the data driven half width criterion found in adaptive_tracking_windows.csv. Output is organized into discovery-set candidate tables, candidate retention audit tables, adaptive tracking window tables, tracked frequency matrices, global correlation measures for each case, dose response tables, and family-wise summary measures. Documentation and metadata files include checksums, citation metadata, DataCite/Schema.org metadata, and CSV schemas. These global and local measures map onto established MAC, FRAC, and FRF-shape localization concepts [3–6].

Secondary analyses can be conducted using two possible modes. An analyst could directly reuse the processed CSV table and figures, without having to read the raw exported data. Alternatively, analysts could use the raw exported data folder (00_raw_exports), and replicate the analysis scripts (04_scripts) to produce the processed output, or even to test different criteria for selection and windows. The package is intended to follow the FAIR data principles [14] and to be cited as a research output in line with the Joint Declaration of Data Citation Principles [15].

## LIMITATIONS

This dataset relates to a single plate configuration, a single realization of boundary conditions, and a single nominal range of preloads, so any conclusions reached using this dataset must be validated using plates with different dimensions and configurations and a wider range of preloads before making any generalizations. The export data relate to frequencies ranging from 1 Hz to 51 200 Hz with 1 Hz frequency increments. Specifically, the high-frequency range beyond 25 kHz consists mostly of the noise floor and does not contain any resonances for this plate, so there is no useful data in this frequency range for FRF correlation.

## ETHICS STATEMENT

The authors have read and follow the ethical requirements for publication. The work does not involve human subjects, animal experiments, or data collected from social media platforms.

## CRediT AUTHOR STATEMENT

**Berkay Kullukcu:** Conceptualization, Methodology, Software, Data curation, Formal analysis, Visualization, Writing – Original draft. **Dina Hannebauer:** Supervision, Resources, Validation, Writing – Review & editing.

## ACKNOWLEDGEMENTS

This work was funded by the European Regional Development Fund (ERDF/EFRE) and the State of Brandenburg within the StaF-Verbund programme, under the project “Systementwicklung für intelligente und automatisierte mobile Inspektion von Schienenfahrzeugen (SIAMIS)”, administered by the Investment Bank of the State of Brandenburg (ILB).

## DECLARATION OF COMPETING INTERESTS

The authors declare that they have no known competing financial interests or personal relationships that could have appeared to influence the work reported in this paper.